\documentclass[aps,amsmath,showpacs,twocolumn,prl,superscriptaddress]{revtex4}

\usepackage{color,amssymb}
\usepackage[dvips]{graphicx}
\usepackage{amsmath,psfrag}
\usepackage{psfrag}
\usepackage{longtable}
\usepackage{dcolumn,epsfig}
\usepackage[normalem]{ulem}
\usepackage{color}

\begin{document}

\newcommand{\comment}[1]{}
\newcommand{\E}{\mathrm{E}}
\newcommand{\Var}{\mathrm{Var}}
\newcommand{\bra}[1]{\langle #1|}
\newcommand{\ket}[1]{|#1\rangle}
\newcommand{\braket}[2]{\langle #1|#2 \rangle}
\newcommand{\be}{\begin{equation}}
\newcommand{\ee}{\end{equation}}
\newcommand{\ba}{\begin{eqnarray}}
\newcommand{\ea}{\end{eqnarray}}
\newcommand{\R}[1]{\textcolor{red}{#1}}
\newcommand{\B}[1]{\textcolor{blue}{#1}}

\title{Universal Quantum Entanglement between an Oscillator and Continuous Fields}

\author{Haixing Miao}
\affiliation{School of Physics, University of Western Australia, WA 6009, Australia}
\author{Stefan Danilishin}
\affiliation{Physics Faculty, Moscow State University, Moscow 119991, Russia}
\affiliation{Max-Planck Institut f\"ur Gravitationsphysik (Albert-Einstein-Institut)
and Leibniz Universit\"at Hannover, Callinstr. 38, 30167 Hannover, Germany}
\author{Yanbei Chen}
\affiliation{Theoretical Astrophysics 130-33, California Institute of Technology,
Pasadena, CA 91125, USA}

\begin{abstract}
Quantum entanglement has been actively sought for in optomechanical and electromechanical
systems. The simplest such system is a mechanical oscillator interacting with a coherent beam,
while the oscillator also suffers from thermal decoherence. For this system, we show that
quantum entanglement is always present between the oscillator and continuous outgoing fields
--- even when the environmental temperature is high and the oscillator is highly classical.
Such universal entanglement is also shown to be able to survive more than one oscillation
cycle if characteristic frequency of the optomechanical interaction is larger than that of
the thermal noise. Furthermore, we derive the effective optical mode that is maximally entangled
with the oscillator, which will be useful for future quantum computing and encoding information
into mechanical degrees of freedom.
\end{abstract}

\maketitle
\noindent{\B{\it Introduction.}} Entanglement, as one of the most fascinating features of
quantum mechanics, lies in the heart of quantum computing and many quantum communication
protocols \cite{BEZ}. Great efforts have been devoted to theoretical and experimental
investigations of quantum entanglements in different systems with discrete or continuous
variables.  Due to advancements in fabricating low-loss optical elements and high-Q mechanical
resonators, the quantum entanglement in optomechanical systems has recently aroused great
interests, especially when many table-top experiments demonstrated significant cooling of
mechanical degrees of freedom via active feedback or passive damping (self-cooling) \cite{Heidmann,Karrai2,
Gigan,Arcizet,Kleckner,Schliesser,Corbitt,Thompson}, which unveils the possibility of
achieving quantum ground state of macroscopic objects \cite{Marquardt,Rae,Genes1}. This
not only paves the way for high-precision measurements but also incorporating mechanical
degrees of freedom as possible medium for storing and retrieving quantum information.

Theoretical analysis shows that by coupling oscillators to a Fabry-Perot cavity, one can
create stationary (Einstein-Podosky-Rosen) EPR-type quantum entanglement between optical
modes and an oscillator \cite{Vitali} or even between two macroscopic oscillators
\cite{Mancini, Hartmann}. In Ref. \cite{helge}, it was shown that entanglement between two oscillators can also be
created by conditioning on continuous measurements of the common and differential optical modes
in a laser interferometer. Interestingly, such entanglement does not depend on the
environmental temperature $T$ explicitly but rather scales as the ratio between characteristic
interaction frequency $\Omega_q$ (equivalent to optical power), and characteristic
thermal-noise frequency $\Omega_F$ (equivalent to $T$).  In contrast, $T$ enters explicitly
and entanglement generally vanishes at high temperature in cases considered in Refs.
\cite{Vitali,Mancini,Hartmann}. This discrepancy arises from the following facts:
(i) The cavity in Ref. \cite{helge} is tuned and thus always stable, while in Refs. \cite{Vitali,Mancini,Hartmann},
stability requirements of the system set an upper limit on $\Omega_q$;
(ii) More importantly, due to finite transmission of the cavity, information leaks into
the environment. Therefore, even regardless of thermal heat bath, the reduced system
consisting of cavity modes and the oscillator is not in a pure state. In Ref. \cite{helge},
however, there can be either no cavity or cavity with very broad bandwidth, the outgoing
fields containing information of oscillator motion are all registered by photodetector.
\begin{figure}
\centerline{\includegraphics[width=0.43\textwidth, bb= 0 0 396 168, clip]{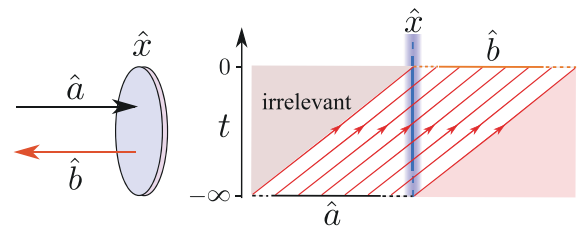}}
\caption{\label{spacetime} A schematic plot of the model and the corresponding spacetime
diagram. Here $\hat x$, $\hat a$ and $\hat b$ denote the oscillator position, ingoing and
outgoing fields respectively. For clarity, we intentionally place $\hat a$ and $\hat b$ on
difference sides of the oscillator world line. The tilted lines represent the light rays.
Up to some instant we are concerned with ($t=0$), the optical fields entering later are
out of causal contact and thus irrelevant.}
\end{figure}

This motivates us to consider that the entanglement scaling obtained in Ref. \cite{helge}
can be inherent in the simplest system --- an oscillator interacting with a coherent beam,
which models the essential process in all above-mentioned optomechanical systems. The model
and its spacetime diagram are shown schematically in Fig. \ref{spacetime}. Similar system
was analyzed previously by Pirandola {\it et al.} \cite{Pirandola}. They used
narrow-detection-band approximation to introduce sideband modes, which maps outgoing fields
into two effective degrees of freedom. In the situation here, sideband modes are not
well-defined, because the interaction turns off at $t=0$ and
only half-space $[-\infty,\,0]$ is involved. Instead, we will directly evaluate the
entanglement between the oscillator and outgoing fields $\hat b$ (infinite degrees of freedom)
using the positivity of partial transpose (PPT) criterion
\cite{Peres,Horodecki, Duan, Simon, Werner, Vidal, Serafini, Adesso}. Only in weak-interaction
and low-thermal-noise limit ($\Omega_q,\,\Omega_F\ll\omega_m$) can we make correspondences
between our results and those obtained in Ref. \cite{Pirandola}.

\noindent{\B{\it Dynamics and Covariance Matrix.}}
The Heisenberg equations for this optomechanical system are simply
\begin{align}\label{1}
\dot {\hat x}(t)&=\hat p(t)/m,\\\label{2}
\dot {\hat p}(t)&=-2\gamma_m\,\hat p(t)-m\,\omega_m^2\hat x(t)+\alpha \,\hat a_1(t)+\hat\xi_{\rm th}(t),\\\label{3}
\hat b_1(t)&=\hat a_1(t),\quad
\hat b_2(t)=\hat a_2(t)+(\alpha/\hbar)\,\hat x(t).
\end{align}
Here $\hat x$ and $\hat p$ are oscillator position and momentum; $\hat a_i$ and $\hat b_i\,(i=1,2)$
are quadratures of ingoing and outgoing optical fields, $\hat a_1\equiv (\hat a+\hat a^{\dag})/\sqrt{2}$
and $\hat a_2\equiv (\hat a-\hat a^{\dag})/(i\sqrt{2})$ (the same for $\hat b_{1,2}$);
$\alpha= (I_0\hbar\,\omega_0/c^2)^{1/2}\equiv (\hbar\,m\,\Omega_q^2)^{1/2}$ is the optomechanical
coupling strength, where $\omega_0$ and $I_0$ denote the laser frequency and optical power respectively
and we have defined $\Omega_q$; $\alpha\,\hat a_1$ is the radiation-pressure term. Since
$[\hat a_1(t),\,\hat a_1(t')]=0$, the presence of thermal noise $\hat \xi_{\rm th}$
ensures the correct commutator between $\hat x(t)$ and $\hat p(t)$ \cite{Gardiner}. The solution to oscillator position $\hat x$ is
\be
\hat x(t)=\mbox{$\int_{-\infty}^{t}$}\,dt'\,G_x(t-t')[\alpha\,\hat a_1(t')+\hat\xi_{\rm th}(t')],
\label{4}
\ee
where Green's function $G_{x}(t)\equiv e^{-\gamma_m t}\sin(\omega_m t)/(m\,\omega_m)$.
The radiation-pressure term $\alpha\, \hat a_1$ induces quantum correlations between the
oscillator and the optical fields, but it is undermined by $\hat \xi_{\rm th}$. The question would be
whether quantum entanglement exists or not after evolving the entire system
from $t=-\infty$ to $0$.
Since variables involved are Gaussian and linear dynamics will preserve
Gaussianity, entanglement is completely encoded in the covariance matrix $\bf V$ of the optomechanical
system. With optical fields
labeled by continuous coordinate $t$, elements of $\bf V$ involving
optical degrees of freedom would be defined in the functional
space ${\cal L}^2[-\infty,\,0]$. Specifically,
\be\label{5}
{\bf V}=
\left[\begin{array}{cc}{\bf A}&{\bf C^{\bf T}}\\{\bf C}&{\bf B}\end{array}\right].
\ee
Here ${\bf A}_{ij}=\langle \vec X_i \,\vec X_j\rangle_{\rm sym}\,(i,j=1,2)$ with vector $\vec X\equiv[\hat x(0),\,\hat p(0)]$ and
$\langle \vec X_i\,\vec X_j\rangle_{\rm sym}\equiv\langle \vec X_i\, \vec X_j+\vec X_j\,\vec X_i\rangle/2$ denoting symmetrized ensemble average;
${\bf C}_{ij}$ and ${\bf B}_{ij}$ should be viewed as vectors and operators in ${\cal L}^2[-\infty,\,0]$.
In the coordinate representation, $(t|\,{\bf C}_{ij})=\langle \vec X_i\,\hat b_j(t)\rangle_{\rm sym}$ and $(t|\,{\bf B}_{ij}\,|t')=\langle \hat b_i(t)\, \hat b_j(t') \rangle_{\rm sym}$, in which $(\,|\,)$ denotes the scalar inner product in ${\cal L}^2[-\infty,\,0]$.

\noindent{\B{\it PPT Criterion.}} According to Refs.~\cite{Werner, Adesso}, in order for one
particle and a joint system of arbitrarily large $N$ particles to be separable, a necessary and
sufficient condition is that partially transposed density matrix $\varrho^{\bf T_1}_{1|N}$
(with respect to the first particle) should be {\it positive semidefinite}, i.e.
$\varrho^{\bf T_1}_{1|N}\ge0$. In the phase space of continuous Gaussian variables, this reduces to
the {\it Uncertainty Principle}
\be
{\bf V}_{\rm pt}+(1/2){\bf  K}\ge0.
\ee
Here commutator matrix ${\bf K}=\bigoplus^{N+1}_{k=1} {2}\,\sigma_y$ with $\sigma_y$ denoting Pauli matrix.
According to the Williamson theorem, there exists a symplectic transformation ${\bf S}\in S_{p(2N+2,\mathbb{R})}$
such that ${\bf S}^{\rm T}{\bf V}_{\rm pt}{\bf S}=\bigoplus^{N+1}_{k=1}{{\rm Diag}[\lambda_k,\,\lambda_k]}$.
Using the fact that
${\bf S}^{\rm T}{\bf K}{\bf S}={\bf K}$, the above {\it Uncertainty Principle} reads
$\lambda_k\ge 1$. If this fails to be the case, i.e. $\exists\,\lambda_k< 1$, the states are entangled.
The amount of entanglement can be quantified by the logarithmic negativity $E_{\cal N}$ \cite{Vidal} and
\be\label{ENdef}
E_{\cal N}\equiv\max[-\mbox{$\sum_{k}$}\ln\lambda_k,\,0]\quad\mbox{for}\,k:\,\lambda_k<1.
\ee

In the case here, $N$ approaches $\infty$. Besides, the partial transpose
is equivalent to time reversal. Therefore ${\bf V}_{\rm pt}={\bf V}|_{\hat p(0)\rightarrow -\hat p(0)}$.
Normalizing $\hat x$ and $\hat p$ with respect to their zero-point values,
the commutator reads $[\hat x, \hat p]=2\,i$. For the optical fields, we set
$[\hat b_1(t), \hat b_2(t')]=2i\,\delta(t-t')$, which gives the coordinate representation of ${\bf K}$.

According to Ref. \cite{Vidal}, $\lambda_k$ can be obtained by solving
\be\label{10}
{\bf V}_{\rm pt}{\bf v}=(1/2)\,\lambda\,{\bf K}\,{\bf v},
\ee
where ${\bf v}\equiv [\alpha_0,\,\beta_0,\,|\alpha),\,|\beta)]^{\bf T}$ with $|f)$ denoting the vector in ${\cal L}^2[-\infty,\,0]$.
Due to uniqueness of $|\alpha)$ and $|\beta)$ in terms of $\alpha_0$ and $\beta_0$ for any $\lambda<1$
(non-singular), Eq. \eqref{10} leads to the following characteristic equation
\begin{equation}\label{det}
\det[{\bf A}+\lambda\,\sigma_y-{\bf C}^{\bf T}(\lambda\,\sigma_y+ {\bf B})^{-1} {\bf C}]
=0
\end{equation}
It can be shown that
\begin{equation}\label{inverse}
(\lambda\,\sigma_y+ {\bf B})^{-1} =
\left[\begin{array}{cc}
1+B_{\lambda}^{\dag}\,M^{-1}B_{\lambda} & -B^{\dag}_{\lambda}\,M^{-1} \\
-M^{-1}B_{\lambda} &M^{-1}\end{array}
\right],
\end{equation}
where we have used the fact that $B_{12}^{\dag}=B_{21}$ in ${\cal L}^2[-\infty,0]$ and have defined
$B_{\lambda}\equiv B_{12}-i\,\lambda$ and $M\equiv B_{22}-B_{\lambda}^{\dag}\,B_{\lambda}$.
The integral operator $M$ can be inverted via {\it Wiener-Hopf method}. Given any function $|g)=M^{-1}|h)$,
in the frequency domain, it reads
\be
\tilde g(\Omega)=\mbox{$\int_{-\infty}^{0}$}dt\,e^{i\,\Omega\,t}M^{-1}|h)=[{1}/{\tilde \psi}_-][{\tilde h}/{\tilde \psi_+}]_-.
\ee
Here $[\;]_-$ means taking the causal part of given function (with poles in lower-half complex plane) and
factorization
\be\label{15}
\tilde\psi_{+}\tilde\psi_{-}\equiv\Lambda+i\lambda(\alpha^2/\hbar)(\tilde G_x-\tilde G_x^*)+(\alpha/\hbar)^2S_F\,\tilde G_x\tilde G_x^{*}
\ee
with $\Lambda\equiv 1-\lambda^2$ and $\tilde G_x$ denoting the Fourier transformation of $G_x(t)$.
In the above equation, $\tilde\psi_{+}$$(\tilde\psi_{-})$
and its inverse are analytic in upper-half (lower-half) complex plane,
$\tilde \psi_{+}(-\Omega)=\tilde \psi^{*}_{+}(\Omega)=\tilde \psi_{-}(\Omega)$. In deriving Eq. \eqref{15}, we have used
$\langle \hat a_i(t)\,\hat a_j(t')\rangle_{\rm sym}=\delta_{ij}\,\delta(t-t')$, and for thermal noise,
Markovian approximation is applied and $\langle\hat\xi_{\rm th}(t)\,\hat\xi_{\rm th}(t')\rangle_{\rm sym}=S_F\,\delta(t-t')$
with $S_F=4\,m\,\gamma_m\,k_B\, T\equiv2\,\hbar\,m\,\Omega_F^2$ and $\Omega_F$ defining the characteristic frequency.
\begin{figure}
\includegraphics[width=0.45\textwidth, bb= 0 0 517 207, clip]{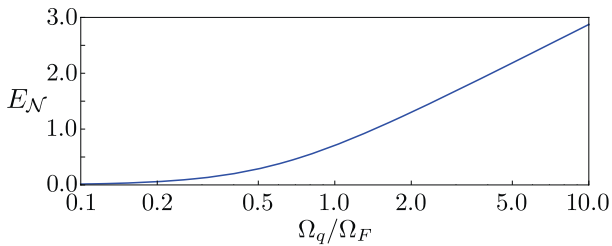}\caption{ Logarithmic negativity
$E_{\cal N}$ as a function of the ratio $\Omega_q/\Omega_F$. A mechanical quality factor $Q_m=10^3$ is chosen. \label{ENuni}}
\end{figure}

\noindent{\B{\it Universal Entanglement.}} Finally, an implicit polynomial equation for the simplectic eigenvalue
$\lambda$ is derived from Eq. \eqref{det}. As it turns out, there always exists one eigenvalue $\lambda$ that
is smaller than one and it only depends on the ratio between $\Omega_q$ and $\Omega_F$, which
clearly indicates the universality of the quantum entanglement.  In Fig. \ref{ENuni},
the corresponding logarithmic negativity (c.f. Eq. \eqref{ENdef}) is shown as a function of $\Omega_q/\Omega_F$.
For a high-Q oscillator $Q_m\equiv \omega_m/(2\gamma_m)\gg 1$,
up to the leading order of $1/Q_m$,  a very elegant expression for $E_{\cal N}$ is derived and it is
\be
E_{\cal N}=({1}/{2})\ln[1+({25}/{8}){\Omega^2_q}/{\Omega_F^2}].
\ee

\noindent{\B{\it Thermal Decoherence.}} To investigate how long such entanglement can
survive under thermal decoherence, after turning off the optomechanical coupling at $t=0$, the mechanical
oscillator freely evolves for a finite duration $\tau$, driven only by thermal noise. Due to thermal decoherence,
entanglement will gradually vanish. Mathematically, the simplectic eigenvalue will become larger than unity
when $\tau$ is larger than the survival time $\tau_s$. By replacing $[\hat x(0), \hat p(0)]$ with
$[\hat x(\tau), \hat p(\tau)]$ and making similar analysis, up to the leading order of $1/Q_m$, $\tau_s$ satisfies
a transcendental equation: $
4\,\Omega_F^4\,\theta_s^2-(2\,\Omega_F^2+\Omega_q^2)^2\sin^2 \theta_s-25\,\omega_m^4=0$,
with $\theta_s\equiv \omega_m\tau_s$. In the case of $\Omega_q< \Omega_F<\omega_m$,
the oscillating term can be neglected, leading to
\be
\theta_s =({5}/{2})({\omega_m}/{\Omega_F})^2={5\,Q_m}/({2\,\bar n_{\rm th}+1}),
\ee
where we have defined the thermal occupation number $\bar n_{\rm th}$ through $k_B T/(\hbar\,\omega_m)=\bar n_{\rm th}+(1/2)$.
For strong interaction $\Omega_q\gg\Omega_F$, the transcendental equation can be solved numerically, showing that
$\theta_s>1$ is always valid.

\noindent{\B{\it Maximally Entangled Mode.}} To gain insights into this entanglement, we apply the techniques in
Ref. \cite{Genes2} and decompose outgoing fields into independent single modes by convoluting them with some
weight functions $f_i$, namely
\be
\hat O_{i}\equiv (f_i|\hat b),~~~~[\hat O_i,\,\hat O_j^{\dag}]=2\,\delta_{ij},
\ee
which requires $(f_i|f_j)=\delta_{ij}$.
If we define $g_{i1}\equiv \Re[f_i]$ and $g_{i2}\equiv \Im[f_i]$, the single-mode quadratures will be
\begin{align}
\hat X_{i}&\equiv (\hat O_i+\hat O_i^{\dag})/\sqrt{2}=\mbox{$\int_{-\infty}^0$}dt\,g_{i1}\,\hat b_1-g_{i2}\,\hat b_2,\label{19}\\
\hat Y_{i}&\equiv (\hat O_i-\hat O_i^{\dag})/(i\sqrt{2})=\mbox{$\int_{-\infty}^0$}dt\,g_{i2}\,\hat b_1+g_{i1}\,\hat b_2.\label{20}
\end{align}

Different choices of weight function will generally give optical modes that have different strength of entanglement with
the mechanical oscillator. The function of particular interest is the one that gives an effective optical mode maximally
entangled with the oscillator. Using the fact that logarithmic negativity is an entanglement monotone, the optimal weight
function can be derived from the following constrained variational equation:
\be\label{var}
({\delta\,E^{\rm sub}_{\cal N}}/{\delta\,g_{i}})+\mu_i\,g_{i}=0 \quad (i=1,2),
\ee
where we have neglected unnecessary indices and $\mu_k$ is Lagrange multiplier due to the constraint $(f|f)=1$ and
$E^{\rm sub}_{\cal N}$ quantifies entanglement in the subsystem consisting of
the oscillator and the effective optical mode $[\hat x(0),\,\hat p(0),\,\hat X,\,\hat Y]$.
As it turns out, the optimal weight functions $g_{1,2}$ have the shape of decay oscillation with poles $\omega$
given by the following polynomial equation
\be\label{pole}
[(\omega-\omega_m)^2+\gamma_m^2][(\omega+\omega_m)^2+\gamma_m^2]+\chi=0,
\ee
where parameter $\chi$ is a functional of $g_{1,2}$ and also depends on $\Omega_q$ and $\Omega_F$. Therefore, the
weight functions are
\be
g_{k}(t)=A_{k}\,e^{\gamma_g\,t} \cos (\omega_g\, t+\theta_{k})\quad(k=1,2),
\ee
with $\gamma_g$ and $\omega_g$ being imaginary and real parts of $\omega$. Analytical solutions to
parameters $A_k, \omega_g, \gamma_g$ and $\theta_{k}$ requires exact expression of $\chi$ in terms
of $g_k$, $\Omega_q$ and $\Omega_F$, which is rather complicated. Instead, we numerically
optimize those parameters to maximize $E^{\rm sub}_{\cal N}$.

Taking into account $(f|f)=1$, $A_1$ and $A_2$ can be reduced to a single parameter $\zeta$,
which is defined through
\be
A_k^2=\frac{4\,\gamma_g(\gamma_g^2+\omega_g^2)\cos^2[\zeta+k(\pi/2)]}{\gamma_g^2+\omega_g^2+\gamma_g^2\cos(2\theta_k)+\gamma_g\,\omega_g\sin(2\theta_k)}.
\ee
From Eq. \eqref{pole}, $\omega_g^2-\gamma_g^2=\omega_m^2-\gamma_m^2$.
In addition, a local unitary transformation (rotation and squeezing) will not change
the simplectic eigenvalue. Without loss of generality, we can fix that $\theta_1=\pi/2$ and $\theta_2=0$.
Therefore, only two parameters $\omega_g$ and $\zeta$ need to be optimized.

\begin{figure}
\includegraphics[width=0.45\textwidth, bb= 0 0 517 152, clip]{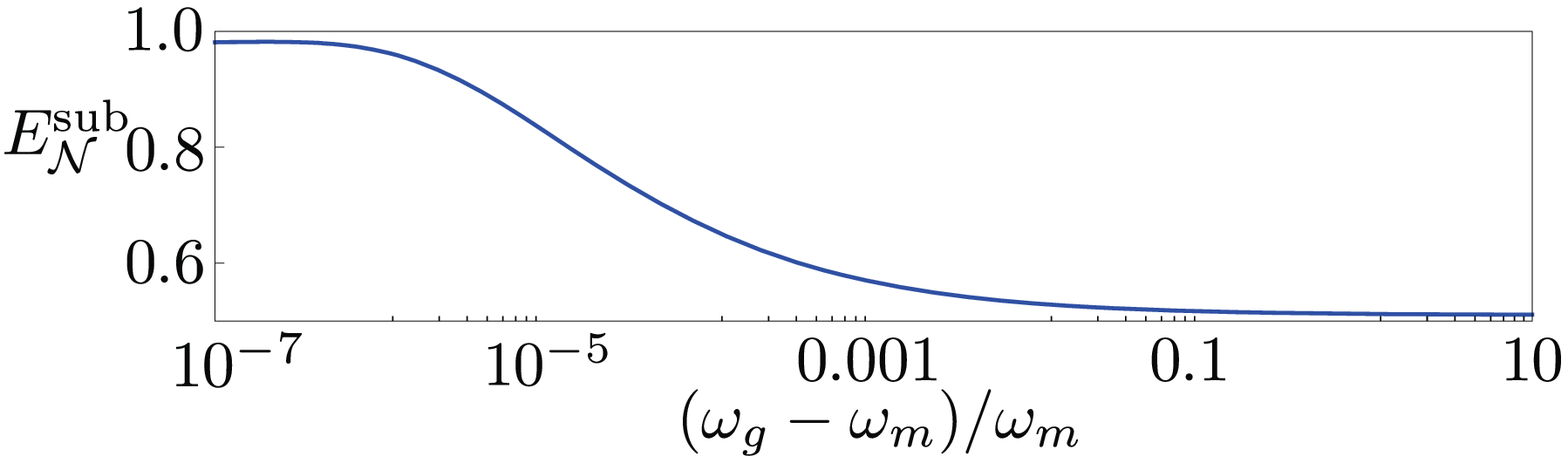}
\includegraphics[width=0.45\textwidth, bb= -5 0 517 152, clip]{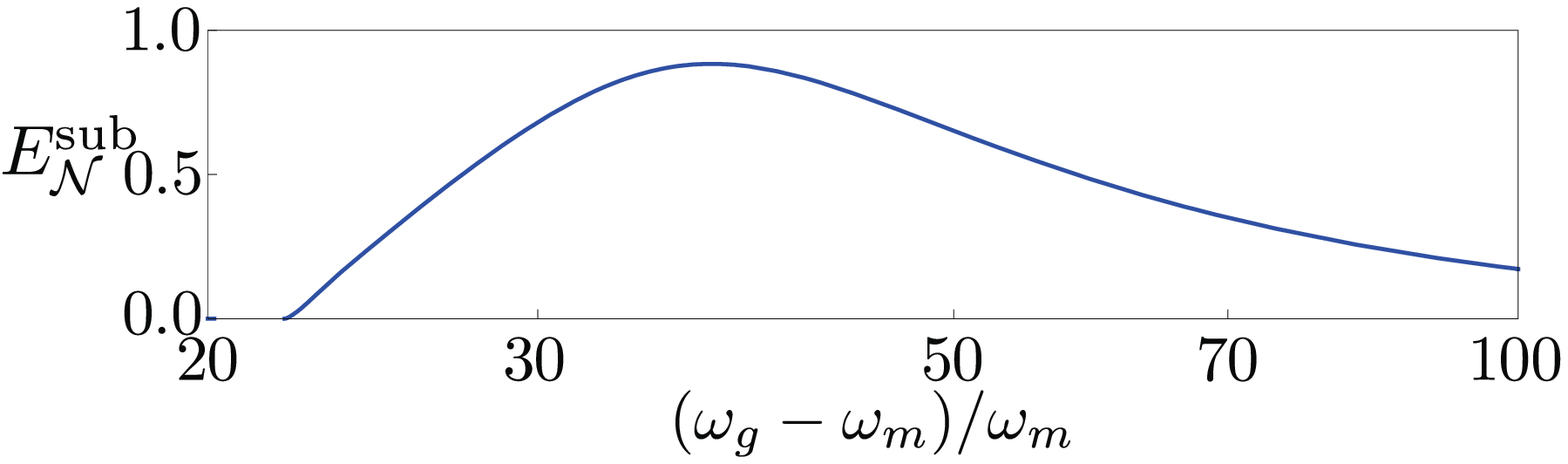}\caption{ Logarithmic negativity
$E^{\rm sub}_{\cal N}$ as a function of quantity $(\omega_g-\omega_m)/\omega_m$ in the
weak-interaction and low-thermal-noise case (upper panel) and strong-interaction and high-thermal-noise case (lower panel).
In the first case, we have chosen $Q_m=10^3,\,\Omega_q/\omega_m=\Omega_F/\omega_m=2\times 10^{-2}$. In the
second case, $Q_m=10^6$ (independent of $Q_m$ for higher $Q_m$), $\Omega_q/\omega_m=50,\,
\Omega_F/\omega_m=20$ and $\zeta=\pi/3.$ \label{EN}}
\end{figure}

In the special case of weak-interaction and low-thermal-noise limit ($\Omega_q,\,\Omega_F\ll\omega_m$),
the optimal $\zeta_{\rm opt}$ is equal to $\pi/4$, which indicates $A_1\approx A_2=2\sqrt{\gamma_m}$
for a high-Q oscillator. Besides, as shown in the upper panel of Fig. \ref{EN}, the optimal $\omega_g^{\rm opt}=\omega_m$, leading to
\be
f(t)=2\sqrt{\gamma_m} e^{\gamma_m t\pm i\,\omega_m t+\phi_0}.
\ee
Therefore, the optimal weight function has the same shape as Stokes and Anti-Stokes sideband modes. This is similar
to what been obtained in Refs. \cite{Pirandola,Genes2}; However, due to causality, the weight function
here is defined in ${\cal L}^2[-\infty,0]$ rather than ${\cal L}^2[-\infty,\infty]$ which is essential for defining
sideband modes.

In the case of strong interaction and high thermal noise ($\Omega_q,\,\Omega_F >\omega_m$), the optimal $\omega_g$
deviates from $\omega_m$ and depends on $\Omega_F$ and $\Omega_q$, as shown in the lower panel of
Fig. \ref{EN}. More generally, the optimal $\zeta_{\rm opt}=\pi/3$ and $\omega_g^{\rm opt}$ can be fitted by $
\omega_g^{\rm opt}\approx(0.64\,\Omega_F^2+0.57\,\Omega_q^2)^{1/2}.$
Correspondingly, the logarithmic negativity can be approximated as
\be
E^{\rm sub}_{\cal N}\approx({1}/{2})\ln[1+({15.\,\Omega_q^2}/({13.\,\Omega_F^2+\Omega_q^2}))],
\ee
which again manifests universality of the entanglement. Therefore, as long as the optimal weight
function is chosen, one can always recover quantum correlations between the oscillator and the outgoing fields.

In principle, by choosing a weight function orthogonal to the optimal one obtained above, one can
derive next-order optimal mode. Repeating this procedure will generate a complete spectrum of effective optical
modes ordered by $E_{\cal N}^{\rm sub}$, which is analogous to obtaining wavefunctions
and corresponding energy levels with variational method.  This not only helps to understand the full
entanglement structure but also sheds light on experimental verifications of such universal entanglement.
Rather than trying to recover the infinite-dimension covariance matrix in Eq. \eqref{5},
we can apply right weight functions to extract different effective optical modes and form low-dimension sub-systems.
Take sub-system consisting of the oscillator and the maximally entangled optical mode for instance, $4\times 4$
covariance matrix can be determined by measuring correlations among different quadratures.
This can be achieved by using a local oscillator with time-dependent phase, which allows to probe both mechanical quadratures
\cite{Miao} and those of the effective optical mode. For expample, a quadrature $\hat O_{\zeta}=\hat X\sin \zeta+\hat Y\cos\zeta$
can be measured with the following local oscillator light:
\be
L(t)\propto L_1(t)\cos\omega_0t+L_2(t)\sin\omega_0t\ee
with $L_1(t)=g_1(t)\cos\zeta+g_2(t)\sin\zeta$ and $L_2(t)=g_2(t)\cos\zeta-g_1(t)\sin\zeta$.
Synthesis of multiple measurements will recover the covariance matrix that we need to verify the
entanglement.

\noindent{\B{\it Conclusion.}} We have demonstrated that quantum entanglement
exists universally in system with a mechanical oscillator coupled to continuous optical fields. The entanglement
measure --- logarithmic negativity displays an elegant scaling which depends on the ratio between characteristic
interaction and thermal-noise frequency.  Such scaling should also apply in electromechanical systems whose
dynamics are similar to what we have considered.

\noindent{\B{\it Acknowledgements.}}
We thank F. Ya. Khalili, H. M\"{u}ller-Ebhardt, H. Rehbein, K. Somiya and our colleagues
at MQM group for invaluable discussions.
Research of H.M. was supported by the Australian Research Council and the Department of Education,
Science and Training. S.D. was supported by Alexander von Humboldt Foundation fellowship.  Y.C.
was supported by the Alexander von Humboldt Foundation's Sofja Kovalevskaja Programme, NSF grants
PHY-0653653 and PHY-0601459, as well as the David and Barbara Groce startup fund at Caltech.

\end{document}